\documentclass[]{article}
\newcommand{\bb}{\begin{eqnarray}}
\newcommand{\ee}{\end{eqnarray}}
\begin{document}
\title{{\bf Patching up the monopole potential}}
\author{Ashok Chatterjee\thanks{e-mail ashok@theory.saha.ernet.in} {} and 
P. Mitra\thanks{e-mail mitra@theory.saha.ernet.in}\\
Saha Institute of Nuclear Physics\\
Block AF, Bidhannagar\\
Calcutta 700 064, INDIA}
\date{hep-th/0312272}
\maketitle
\begin{abstract}
It is well known that a vector potential cannot be defined over
the whole surface of a sphere around a magnetic monopole. A recent claim
to the contrary is shown to have problems. It is explained however that a  
potential of the proposed type works if two patches are used instead of one. 
A general derivation of the Dirac quantization condition attempted
with a single patch is corrected by introducing two patches. 
Further, the case of more than 
two patches using the original Wu-Yang type of potential is discussed in brief.
\end{abstract}

\bigskip

While magnetic monopoles have not been seen experimentally,
they have continued to be of interest to theoreticians.
In standard electrodynamics, magnetic monopoles are not admitted,
and the magnetic field has zero divergence. If a magnetic monopole
is present at a point, the divergence is a delta function with support
at that point, and it
is not possible to introduce a vector potential 
which is nonsingular everywhere.
Dirac\cite{dirac} found that there is at least a string of singularities
from that point to infinity,
though the location of the string is arbitrary, like a branch cut.
Instead of working with a string, Wu and Yang \cite{WY} showed that it may be
more convenient to work with two different nonsingular 
potentials on two overlapping
patches, with a gauge transformation connecting them in the overlap.
In both approaches the quantum theoretic description of an electrically 
charged particle requires the quantization of its charge, 
if a monopole is assumed to exist.

Recently there has been a claim \cite{gp} about the construction of a
nonsingular vector potential for the monopole field using a {\it single} patch.
This could only be possible if the basis of the Wu-Yang formalism were
incorrect, and indeed that is what has been asserted. One of our aims is 
to reexamine this question and 
to show that the {\it single}-patch potential does not work. 
The use of two patches, with a potential in the $\phi$ direction \cite{WY}, 
is standard. 
We show how the vector potential of \cite{gp}, which is in the
$\theta$ direction, can be extended to two
patches, where it makes sense. 
A general argument for the Dirac quantization, presented for a single
patch in \cite{gp}, also does not work. We show how it can be
corrected for two patches.   We then
discuss the possibility of using the $\phi$-directional potential of
\cite{WY} in {\it more than two} patches: although two
patches are enough, one can use more, contrary to claims in
\cite{gp}. There arise interesting issues
of consistency in this situation, but the standard quantization
condition continues to hold.   

The potential proposed \cite{gp} for a monopole of strength $g$ at 
the origin is locally given by the 1-form 
\bb
A=-g f(\phi)  \sin\theta d\theta.
\ee
This leads to the 2-form field
\bb
F=dA={g\over r^2} f'(\phi) rd\theta \wedge r\sin\theta d\phi,
\ee
which reduces to the standard expression for the magnetic field for 
the monopole at the origin if and only if $f'$ is unity, {\it i.e.,}
\bb
f(\phi)=\phi+{\rm constant}.
\ee
This choice makes $f$ multiple valued in
the sense that $\phi$ and $\phi +2\pi$ give different values. If the angle
were taken mod $2\pi$, the derivative would no longer be unity: 
the mod involves a sawtooth function, whose derivative is
a sum of delta functions with supports at integral values of $\phi/2\pi$,
so that the magnetic field would
have singularities there. If $\phi$ is restricted to an
interval of size 2$\pi$, the potential becomes discontinuous
at the boundaries of the interval. In general, for any $f$, 
one may note that the line integral on a closed
curve C with $\theta=$ constant on the sphere $r=$ constant is well defined
and given by
\bb
\oint_{\rm C} A=0.
\ee
If Stokes' theorem, which is valid for a nonsingular potential, is applicable, 
it implies that the flux through a surface bounded by C is zero. This is
obviously not the case for the magnetic field due to a monopole.
Therefore, the potential cannot be nonsingular. 
It should be pointed out that the line integral has been proposed
\cite{gp} to be replaced by an integral over a complicated path so that
it ceases to be zero, but such a replacement can be justified only
if the potential is defined on two patches -- as discussed below -- instead
of one. Otherwise it is {\it ad hoc} to replace the well-defined line integral
by something which agrees with the answer expected from Stokes' theorem.
  
In fact, the impossibility of constructing a vector potential 
using a single patch is a theorem in \cite{WY}. The idea is that the
flux through the part of the sphere $r=$ constant with $\theta<\theta_0$,
where $0<\theta_0<\pi$, increases as $\theta_0$ increases from zero to
$\pi$, reaching the total flux $4\pi g$ in the limit. However,
for each $\theta_0$, the flux is also equal to $\oint_{\rm C[\theta_0]} A$,
where ${\rm C[\theta_0]}$ is the curve $\theta=\theta_0$ on the sphere,
if $A$ can be defined over the whole sphere. In the limit $\theta_0\to\pi$,
this curve reduces to a point, 
so that the line integral of a nonsingular $A$ must vanish,
yielding a contradiction $4\pi g\stackrel{?}{=}0$ for $g\neq0$. 

This contradiction was dismissed in \cite{gp},
on the ground that the 
curve ${\rm C[\theta_0]}$ is also the boundary of another part of the
spherical surface, namely that with $\theta>\theta_0$, 
which vanishes in the limit $\theta_0\to\pi$,
so that the flux through it is zero.
But this is a red herring. For a potential which is regular everywhere,
the line integral on the curve
must be equal to the surface integral on {\it each} of the two parts (taken 
in the appropriate sense). The
contradiction between the flux on the whole sphere and the
line integral on a curve of vanishing length can be recast as a
mismatch between the fluxes on the whole surface and its null complement.
This contradiction cannot be removed without removing the monopole
if one persists in the attempt to use a single patch.
One needs at least two patches to construct a nonsingular
$A$ on a sphere $r=$ constant.
 
As is well known, one
possibility \cite{WY} is to take the two patches as $\theta<\theta_1$ and
$\theta>\theta_2$ with $0<\theta_2<\theta_1<\pi$. If the line integral
on a curve with constant $\theta$ and varying $\phi$ is to be nonzero,
the potential 1-form must have a $d\phi$ piece. Equating the line integral
to the surface integral in the first patch, one finds this piece to be
\bb
A^{(1)}=g(1-\cos\theta)d\phi \quad 0<\theta<\theta_1.\label{A1}
\ee
Similarly the surface integral appropriate for the second patch yields
\bb
A^{(2)}=g(-1-\cos\theta)d\phi \quad \theta_2<\theta<\pi.
\ee
In the overlapping region, where both of these expressions are valid,
they are clearly unequal, showing the impossibility of having a single patch,
but their difference is closed, and as expected from Poincar\`{e}'s lemma, 
locally exact:
\bb
A^{(1)}-A^{(2)}=d\chi,\label{chi}
\ee
with
\bb
\chi=2g\phi.
\ee
Although $\chi$ is multiple-valued, $d\chi$ is well defined, so there is no
problem in the context of the classical theory. However, if there
is a particle with electric charge $e$ in this magnetic field, and
it is described by a quantum mechanical wavefunction, the gauge
transformation of this wavefunction involves the factor $\exp (ie\chi
/\hbar c)$ and this has to be single-valued. This requires
\bb
2ge=n\hbar c \quad n~ {\rm integral},\label{Dirac}
\ee
which is the Dirac quantization condition. We have reviewed this
well known analysis in some detail because we want to extend it in
two ways: to the potential proposed in \cite{gp} and to the case
of more than two patches which has also been questioned.

If we seek to use the kind of potential proposed in \cite{gp},
we have to take patches in the $\phi$ direction rather than
the $\theta$ direction. Thus, using two patches as before, we may set
\bb
A^{(1)}=-g \phi  \sin\theta d\theta \quad \phi_1<\phi<\phi_1',
\ee
and
\bb
A^{(2)}=-g \phi  \sin\theta d\theta \quad \phi_2<\phi<\phi_2',
\ee
with
\bb
\phi_2<0<\phi_1<\phi_2'<\phi_2+2\pi<\phi_1'<2\pi.
\ee
The two patches together cover the full region $(0,2\pi)$ with
overlaps $R_1\equiv(\phi_1,\phi_2')$ and 
$R_2\equiv(\phi_2+2\pi,\phi_1')$. 
In these overlapping regions, the two potentials are gauge related:
\bb
A^{(1)}-A^{(2)}=\cases{0 & $\phi$  in $R_1$\cr
2\pi gd\cos\theta & $\phi$ in $R_2$}.
\ee
The $2\pi$ comes from the fact in $R_2$, the coordinate $\phi$ 
has values differing by $2\pi$ in the two patches.
One may write these in the form (\ref{chi}) with
\bb
\chi=\cases{0& $\phi$  in $R_1$\cr
2\pi g(\cos\theta+p)& $\phi$ in $R_2$},
\ee
where $p$ is a constant.
For $\theta$ different from $0,\pi$, the regions $R_1,R_2$ are 
non-overlapping. However, when $\theta$
goes to either of the limiting values, 
$\phi$ is undefined, and the two regions cannot be distinguished. Clearly, 
the two expressions for $\chi$ cannot
be made to agree at both $\theta=0$ and $\theta=\pi$ by any choice of
$p$, but by choosing it to be equal to either -1 or +1, they
can be made to agree at either $\theta=0$ or $\theta=\pi$. At the other
point, $\chi$ has two unequal values, but as in the earlier discussion,
the gauge transformation $\exp(ie\chi/\hbar c)$ of the
wavefunction of a particle with electric charge $e$ becomes single-valued 
if and only if the Dirac 
condition (\ref{Dirac}) holds. In this way, the $\theta$-directional potential
\cite{gp} can be made to work with two patches, and the Dirac condition
emerges as the consistency condition. The evaluation of  a line integral
on a loop with constant $\theta$ can be carried out now in the manner
indicated in \cite{gp}, the deformation of path suggested there being
equivalent to the necessary gauge transitions between the two patches.

An attempt was made in \cite{gp} to derive the Dirac condition without
invoking any specific form of the potential, but it is based on a single patch,
which as we know, is impossible. However, it is not difficult to repair the
derivation using two patches, which we shall now do. Suppose the sphere
around the monopole is covered by two overlapping patches 
and nonsingular potentials
$A^{(1)},A^{(2)}$ defined on them to describe the monopole magnetic field.
Then in the overlapping region,
\bb
F=dA^{(1)}=dA^{(2)}.
\ee
Thus in this region the two potentials must again be related by 
(\ref{chi}), with a $\chi$ determined by $A^{(1)},A^{(2)}$.
Now consider a closed loop $C$ which lies entirely in the
overlap region and is non-contractible within the overlap. Then the two
surfaces $S_1,S_2$ bounded by $C$ lie in the two patches and cover the
full sphere. The application of Stokes' theorem to the two patches
leads to the two separate equations
\bb
\oint_C A^{(1)}&=&\int_{S_1}F,\nonumber\\
\oint_C A^{(2)}&=&-\int_{S_2}F,
\ee
where the difference in orientation of the two surfaces relative
to a given direction of traversal of $C$ has been taken into account.
The above two equations, on subtraction, lead to the result that
\bb
\Delta\chi\equiv\oint_Cd\chi=\int_{S_1\cup S_2}F=4\pi g\neq0.
\ee
This shows that the gauge transformation function $\chi$ is not single-valued,
but the relevant factor for the wavefunction of a particle of charge $e$ is
$\exp(ie\Delta\chi/\hbar c)$ and it is
single-valued if and only if the Dirac condition (\ref{Dirac}) is satisfied.
This demonstration does not rely on explicit formulas for $A$ in the $\phi$ or
$\theta$ directions. But it does involve $A$ defined on two patches. A gauge
invariant derivation without introducing $A$ was given in \cite{jackiw}
on the basis of the assumption that translation operators are associative.

Finally we turn to the case with more than two patches.
The extension of the analysis of \cite{WY} using two
patches to the case of more patches has been questioned \cite{gp}. 
A clarification is in order here.
If there are $N$ patches ($N$  integral),
the first patch has the same potential (\ref{A1}) as before,
and in the last patch one has 
\bb
A^{(N)}=g(-1-\cos\theta)d\phi \quad \theta_N<\theta<\pi.
\ee
but now there are other patches in between, and for these patches,
the potentials are of the form
\bb
A^{(r)}=g(a_r-\cos\theta)d\phi \quad \theta_r<\theta<\theta_{r'},
\ee
where the angles $\theta_r$ and $\theta_r'$ for different $r$ are such
that the successive patches overlap. The
constants $a_r$ can be fixed only for $r=1,N$ by requiring regularity
at 0 and $\pi$ for the potentials on the two 
extreme patches: 
\bb
a_1=1,\quad a_N=-1.
\ee 
Classically, once again, there is no problem with
gauge equivalence of these expressions in overlapping regions, but
if there is an electrically charged particle being treated quantum
mechanically, one needs at least the conditions
\bb
(a_r-a_{r+1})ge=n_r\hbar c \quad n_r~ {\rm integral}
\label{sew}\ee
from successive overlapping patches, as in (\ref{Dirac}). 
If there are further overlaps,
for example between the $r$th and the $s$th with $|r-s|\neq 1$, there
may apparently be further conditions, but they are not independent
and can be obtained from those in (\ref{sew}). 
In any case, by adding up the conditions
indicated above, and using the values of $a_1,a_N$, one gets the
Dirac condition (\ref{Dirac}) again, with
\bb
n=(n_1+...+n_{N-1}).
\ee
While this is a straightforward 
generalization of the case with two patches, there is something
new here: the $a_r$ for $2\leq r\leq N-1$ are not determined by
regularity conditions. They are not totally arbitrary, however. 
In fact, the allowed values are the rational numbers
\bb
a_r=1-{2(n_1+...+n_{r-1})\over (n_1+...+n_{N-1})},
\ee
as follows from the overlap conditions. 
We add however that the number and allowed values of $a_r$ are dictated
by our choice of patches and the physics of the problem is entirely
contained in the Dirac condition (\ref{Dirac}).

To summarize, we have shown that the single-patch potential claimed
to describe a magnetic monopole does not work. The standard proof that at
least two patches are needed is valid. 
A nonsingular version of the $\theta$-directional 
potential using two patches has been
worked out. A general derivation of the Dirac condition using two patches 
has been presented following the earlier single-patch attempt. 
More than two patches can also be
used, and we have worked out the possible forms of the potential in the
different patches using the more familiar $\phi$-directional 
potential \cite{WY}.

\section*{Acknowledgment}
We would like to thank our colleague Palash Baran Pal for many
discussions on \cite{gp}.
We are also grateful to Amit Ghosh and
Parthasarathi Majumdar for several conversations on the subject.

\end{document}